\documentclass[11pt]{article}
\usepackage{moriond}
\usepackage{epsfig}

\def\NIMA{{\em Nucl. Instrum. Methods} A}

\def\PLB{{\em Phys. Lett.}  B}

\def\PRD{{\em Phys. Rev.} D}

\def\be{\begin{equation}}
\def\ee{\end{equation}}
\def\bea{\begin{eqnarray}}
\def\eea{\end{eqnarray}}

\newcommand{\Photo}{\includegraphics[height=35mm]{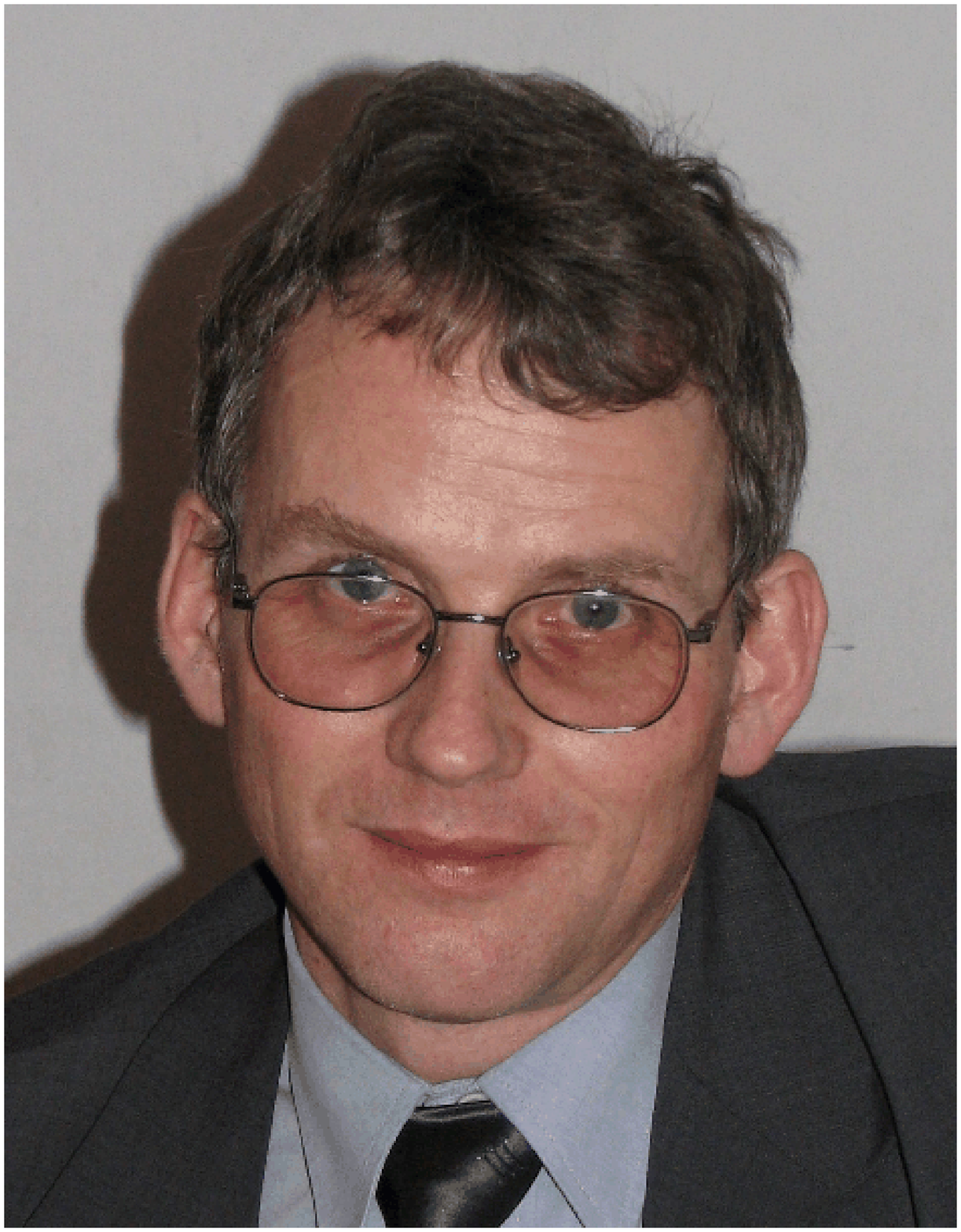}}

\begin{document}
\vspace*{4cm}
\title{ELECTROWEAK RESULTS FROM HERA}

\author{ A.F. \.ZARNECKI \\
  (on behalf of the H1 and ZEUS collaborations)}

\address{Faculty of Physics, University of Warsaw,    
              Ho\.za 69, 00-681 Warszawa, Poland}

\maketitle\abstracts{
Neutral and charged current deep inelastic $ep$ scattering with
longitudinally polarised lepton beams has been studied with the H1
and ZEUS detectors at HERA. The differential cross sections were
measured in the range of four-momentum transfer squared, $Q^{2}$, up to
50'000 GeV$^{2}$, where electroweak effects become clearly visible. 
The measurements were used to determine the structure function $xF_{3}$ 
and to constrain vector and the axial-vector couplings of the light quarks 
to the $Z^{0}$ boson. The polarisation dependence of the charged current
total cross section was also measured. \\
Limits on flavour changing neutral current processes were computed
from the search for single-top production. The elastic $Z^{0}$ production
cross section was measured to be in agreement with the SM prediction. 
Limits on new physics phenomena at high $Q^{2}$ were also derived within the
general framework of four-fermion $eeqq$ contact interactions.
}


\section{Introduction}

The HERA accelerator was built at DESY, Hamburg,  to study 
electron-proton and positron-proton collisions at center
of mass energies of up to 320 GeV.
Scattering events were reconstructed in 
two multi-purpose detectors, H1\cite{h1det} and ZEUS\cite{zeusdet},
both equipped with silicon tracking, drift chambers, hermetic
calorimetry and muon detector systems.  
During the so called HERA~I running phase (1994-2000), about 100 $pb^{-1}$
of data were collected per experiment, mainly coming from $e^+ p$ collisions.
After the collider upgrade in 2000-2001, resulting in significant
increase of luminosity, about 400 $pb^{-1}$ of data  per experiment
were collected in the so called HERA~II phase (2002-2007).
Moreover, spin rotators installed at the H1 and ZEUS interaction
regions allowed the operation with longitudinal electron or positron
polarisation.
With an average lepton beam polarisation of about 30-40\% and a significant
increase of integrated luminosity (especially for the
$e^- p$ sample), HERA~II significantly extended the physics reach
of the experiments.
Different detector configurations and
complementary event reconstruction methods used by two collaborations
allowed for an additional reduction of not only the statistical but also
the systematic uncertainties in the combined analysis of H1 and ZEUS data.


\section{Deep Inelastic $e^{\pm}p$ Scattering}

\begin{figure}[t]
\centerline{\epsfig{file=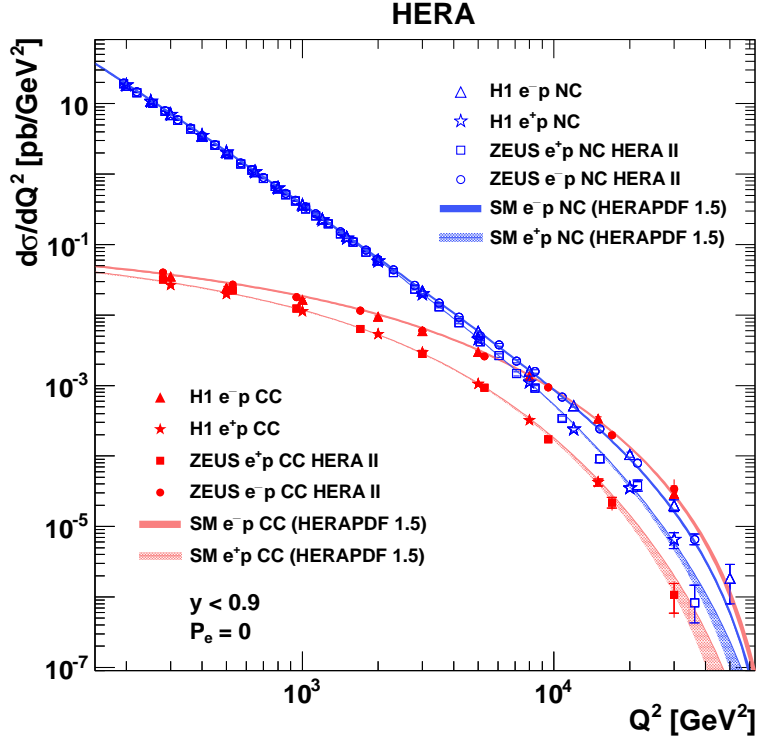,width=0.65\textwidth}}
  \caption{
$Q^{2}$ dependence of the NC and CC cross sections $d\sigma/dQ^{2}$
for the combined HERA~I+II unpolarised $e^{-}p$ and $e^{+}p$ data. 
The inner and outer error bars represent the statistical
and total errors, respectively.
The H1 and ZEUS data are compared to
the Standard Model expectation based on the HERAPDF~1.5
parametrisation. 
    \label{fig:nccc} }
\end{figure}

Deep Inelastic Scattering (DIS) was the main process studied at HERA.
Precise HERA data are in excellent agreement with Standard Model
predictions over many orders of magnitude, as illustrated in
Figure~\ref{fig:nccc}. 
At very high squared momentum transfers $Q^2$, comparable with masses
of the $W$ and $Z$ bosons squared, the contributions from neutral current
(NC) and charge current (CC) processes become comparable in size,
which is a clear demonstration of electroweak unification. 

High $Q^2$ NC DIS cross section, neglecting  radiative corrections,
can be written in terms of three generalised structure functions
 $\tilde{F}_{2}$, $x\tilde{F}_{3}$ and $\tilde{F}_{L}$: 
\begin{eqnarray*}
\frac{d^2 \sigma ^{\rm NC} (e^\pm p)}{dx dQ^2} & = & 
\frac{2\pi \alpha ^2}{xQ^4} 
\left[ Y_{+}\tilde{F}_{2}^{\pm} \mp Y_{-}x\tilde{F}_{3}^{\pm} 
      -y^{2} \tilde{F}_{L}^{\pm} \right] 
\end{eqnarray*}
where: $Y_{\pm}=1\pm(1-y)^{2}$.
The sensitivity to electroweak effects is mainly due to the interference
of photon and $Z$ boson exchange which dominates over pure $Z$
exchange effects in most of the kinematic range covered at HERA. 
Corresponding contributions to the generalised structure functions can
be written as:
\begin{eqnarray*}
\tilde{F_2}^\pm & = &  
         F_2^{\gamma} - (v_e \pm P_e a_e) \chi_{Z} \; F_2^{\gamma Z} \;\; + \;
        (v_e^2 + a_e^2 \pm 2 P_e v_e a_e) {\chi_{Z}^{2}} \;\; F_2^{Z} \\[4mm]
x\tilde{F_3}^\pm & = &  
      \hspace*{5mm} - \; (a_e \pm P_e v_e) \chi_{Z} \; xF_3^{\gamma Z}
       + (2 v_e a_e \pm P_e(v_e^2 + a_e^2)) {\chi_{Z}^{2}} \; xF_3^{Z}
\end{eqnarray*}
where $P_{e}$ is the lepton beam polarisation and 
$ \chi_{Z}  =  \frac{1}{\sin ^{2} 2\theta_W} 
(\frac{Q^{2}}{M_{Z}^{2}+Q^{2}})$. 
Access to electroweak effects is provided by measuring differences between 
cross section for different charges and polarisation, thereby removing 
the pure photon-exchange part, described by $F_{2}^{\gamma}$. 
In particular,  polarisation asymmetries
can be used to constrain contribution from $F_{2}^{\gamma Z}$ structure
function, which is sensitive to $d/u$ ratio at high-$x$ and to the
vector quark couplings, $v_{q}$.  
A comparison of $e^-p$ and $e^+p$ cross sections, i.e. the measurement
of the charge asymmetry, accesses the $xF_{3}^{\gamma Z}$ contribution, 
which is dominated by the valence quark distributions at high $Q^2$ 
and is sensitive to the axial-vector quark coupling,~$a_{q}$.

\begin{figure}[t]
\centerline{\epsfig{file=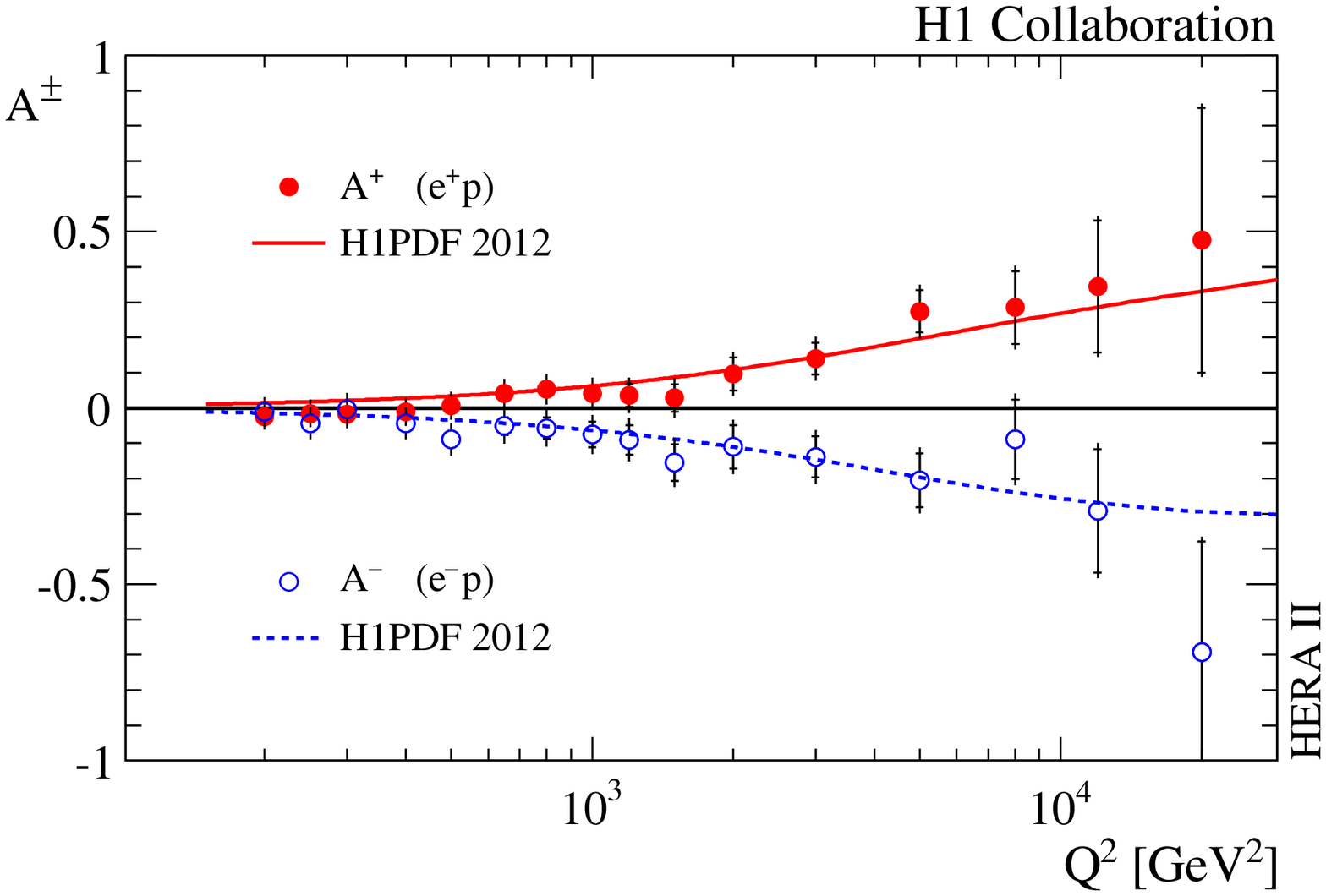,width=0.5\textwidth} 
            \epsfig{file=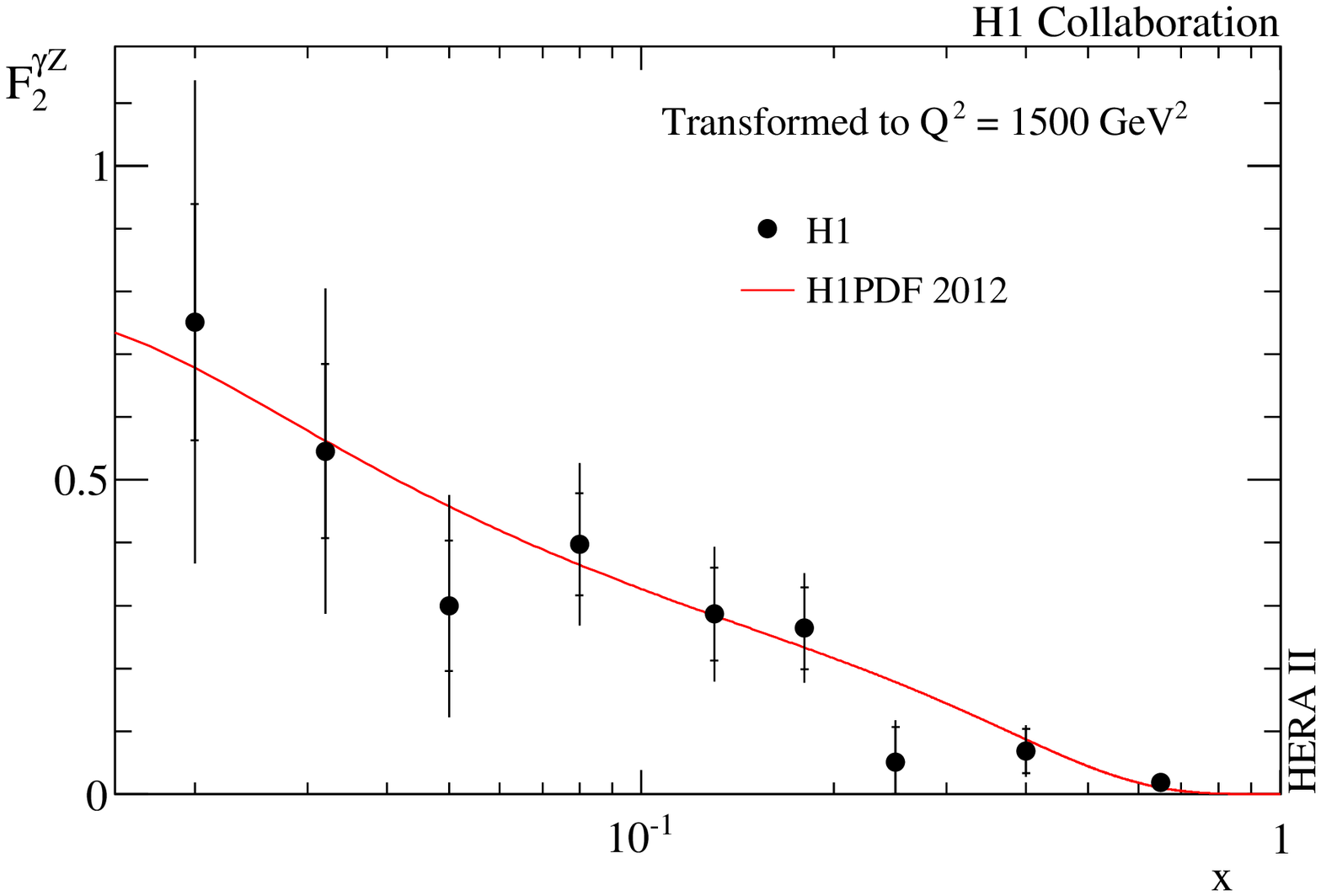,width=0.5\textwidth}}
  \caption{
Left: 
$Q^{2}$ dependence of the polarisation asymmetry $A^{\pm}$, as
measured by the H1 Collaboration, compared to the Standard Model
expectation. 
Right:
structure function $F_{2}^{\gamma Z}$, extrapolated to $Q^{2}$ = 
1~500~GeV$^{2}$, for the H1 data (solid points) and the expectation
from H1PDF 2012 (solid curve). The inner error bars represent the
statistical uncertainties and the full error bars correspond to the
total uncertainties. 
    \label{fig:ncpol} }
\end{figure}

\begin{figure}[t]
\centerline{\epsfig{file=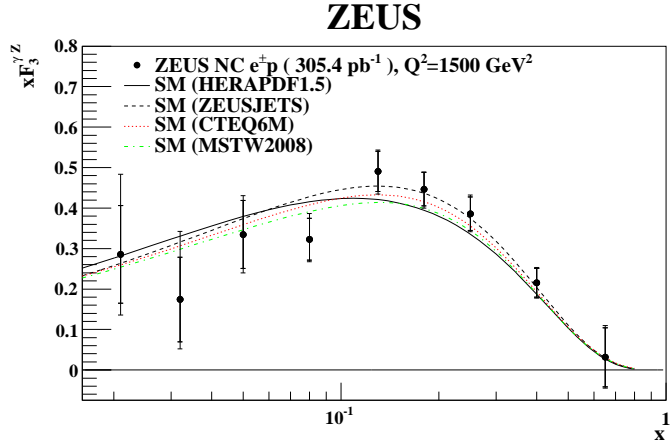,width=0.6\textwidth}}
  \caption{
The structure function $xF_3^{\gamma Z}$, extrapolated to 
$Q^{2}$ = 1~500~GeV$^{2}$, and plotted as a function of x.
The closed circles represent the ZEUS data, with the inner error bars
showing the statistical uncertainties while the outer 
ones show the statistical and systematic uncertainties added in
quadrature. The curves show the predictions of the SM evaluated using 
 different PDF parametrisations, as indicated in the plot. 
    \label{fig:f3} }
\end{figure}

Both, the H1 and ZEUS experiments, have measured neutral current DIS cross
sections for both charges and both helicity states.
Results on polarised cross-section asymmetries,
$A^{\pm}$, as obtained by the H1 Collaboration\cite{desy-12-107},
are shown in Figure~\ref{fig:ncpol} (left).
Parity violation due to $\gamma - Z$ interference is clearly visible,
in agreement with  Standard Model expectations.
The measurement can also be used to extract the $F_{2}^{\gamma Z}$
contribution, as illustrated in Figure~\ref{fig:ncpol} (right).
Shown in Figure~\ref{fig:f3} is the measurement of the structure
function $xF_3^{\gamma Z}$ by the ZEUS Collaboration\cite{desy-12-145};
a similar measurement has been also performed by H1 \cite{desy-12-107}.
The results agree very well with the Standard Model predictions 
obtained from NLO QCD fits to inclusive data.

\begin{figure}[tp]
\begin{center} 
  \epsfig{file=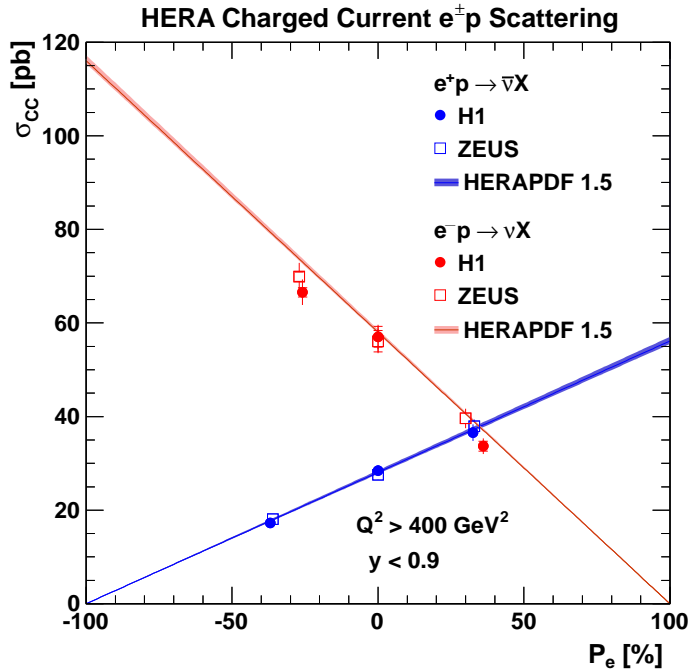,width=0.6\textwidth}  
\end{center}
  \caption{
Dependence of the total $e^{\pm}p$ CC cross sections on the longitudinal
lepton beam polarisation, $P_e$. The H1 and ZEUS data are compared to
the Standard Model expectations based on the HERAPDF~1.5
parametrisation (shaded bands). 
    \label{fig:cctot} }
\end{figure}

The CC DIS  cross-section dependence on the longitudinal
lepton-beam polarisation also follows exactly the Standard Model
predictions, as shown in Figure~\ref{fig:cctot}. For electron beams,
only the left-handed initial state contributes to the scattering 
cross-section, whereas for positrons only the right-handed state contributes.
This is expected from the chiral structure of the model. 
The results can be used to set limits on the possible contribution from
right-handed currents. Assuming SM couplings and a light right
handed $\nu_{e}$, the H1 collaboration excluded the existence of right
handed weak bosons of masses below 214 (194) GeV for $e^{-}p$ ($e^{+}p$)
scattering.

\begin{figure}[tp]
\epsfig{file=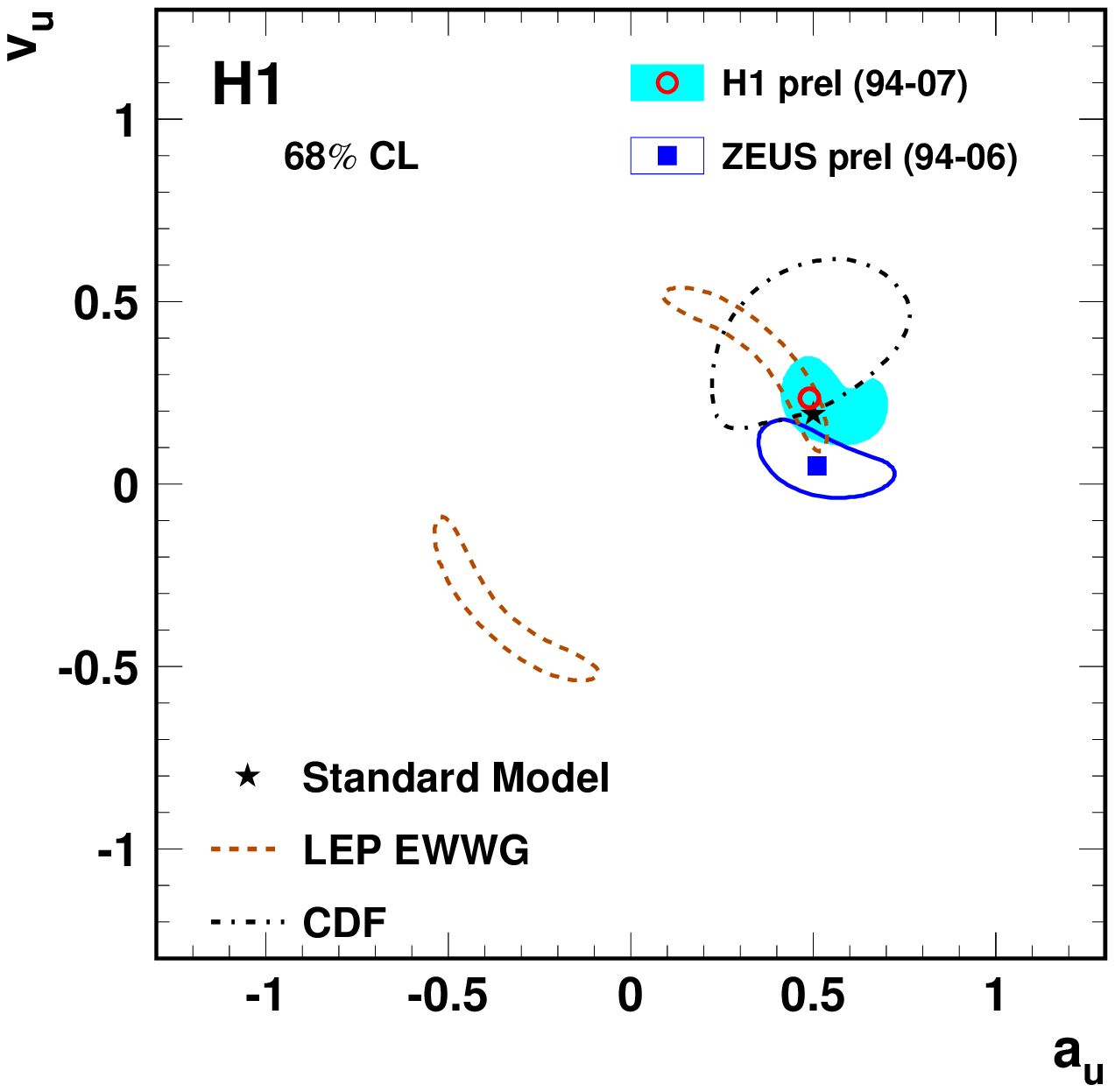,width=0.5\textwidth}
\epsfig{file=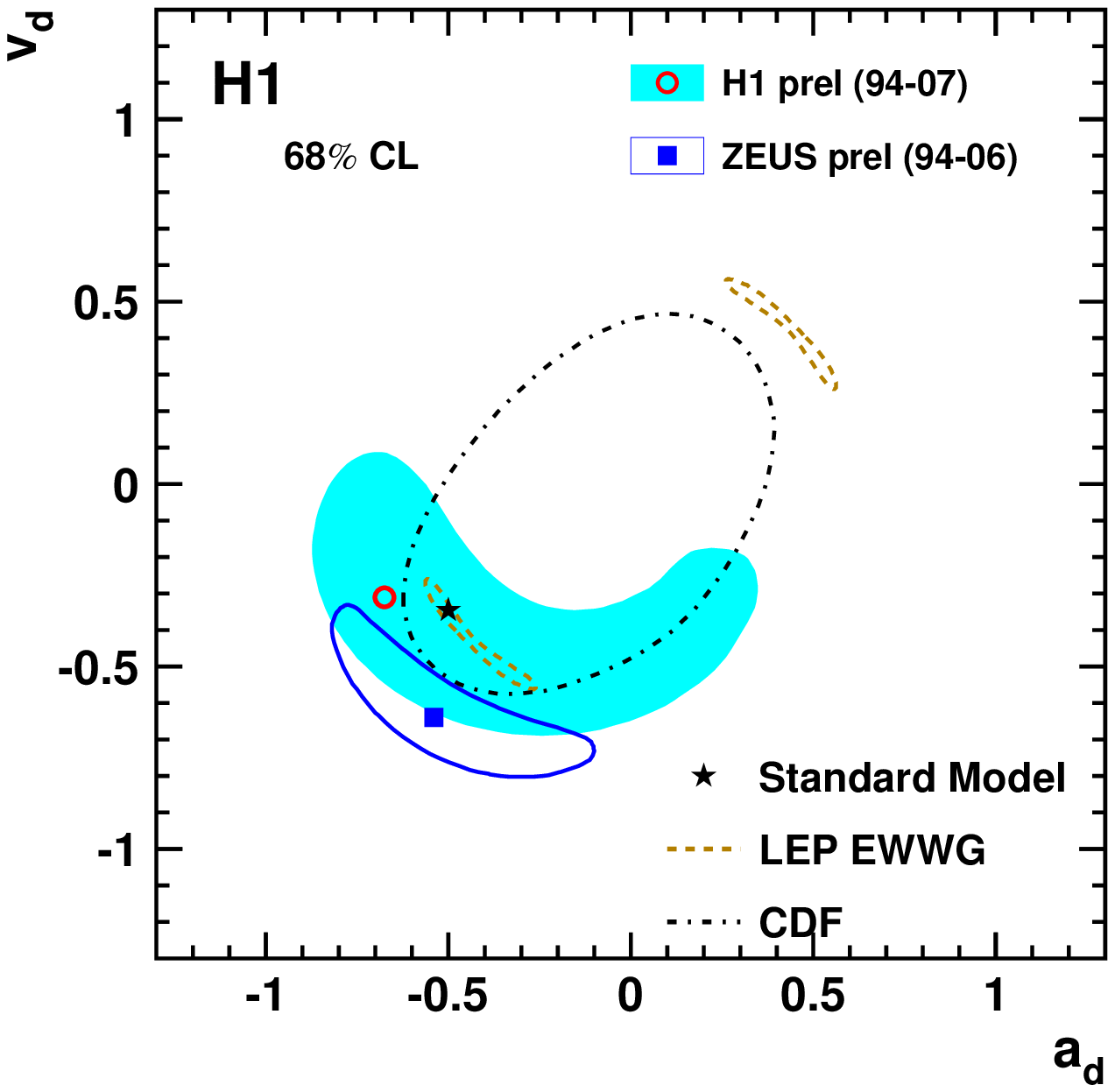,width=0.5\textwidth}
  \caption{
Results at 68\% C.L. on the weak neutral current couplings 
of u (left plot) and d (right plot) quarks to the $Z^0$ boson, determined 
from an H1 analysis of HERA I and HERA II data. Results from a ZEUS analysis
and limits determined by the CDF experiment and the LEP Electroweak Working 
Group (open contours) are included for comparison. The stars show the
expected SM values. 
    \label{fig:ewfit} }
\end{figure}

The wide kinematic range covered, and the high precision of the NC and CC DIS
measurements for polarised $e^{-}p$ and $e^{+}p$,
allow not only the determination of the parton distribution functions 
of the proton from the HERA data alone, but also allow the simultaneous 
determination of PDFs and electroweak parameters.
Results on the weak neutral current couplings of $u$ and $d$ quarks to
the $Z^0$ boson, as obtained by the H1
Collaboration\cite{h1prelim-10-042}, are presented in 
Figure~\ref{fig:ewfit}.  
Results from the earlier ZEUS analysis
and limits determined by the CDF experiment and the LEP EWWG
are included for comparison. 
A good agreement with the Standard Model predictions is observed.
Determinations of the light-quark couplings at HERA turn out
to be competitive in precision with those obtained from the Tevatron and LEP
experiments. 

\begin{figure}[tp]
\epsfig{file=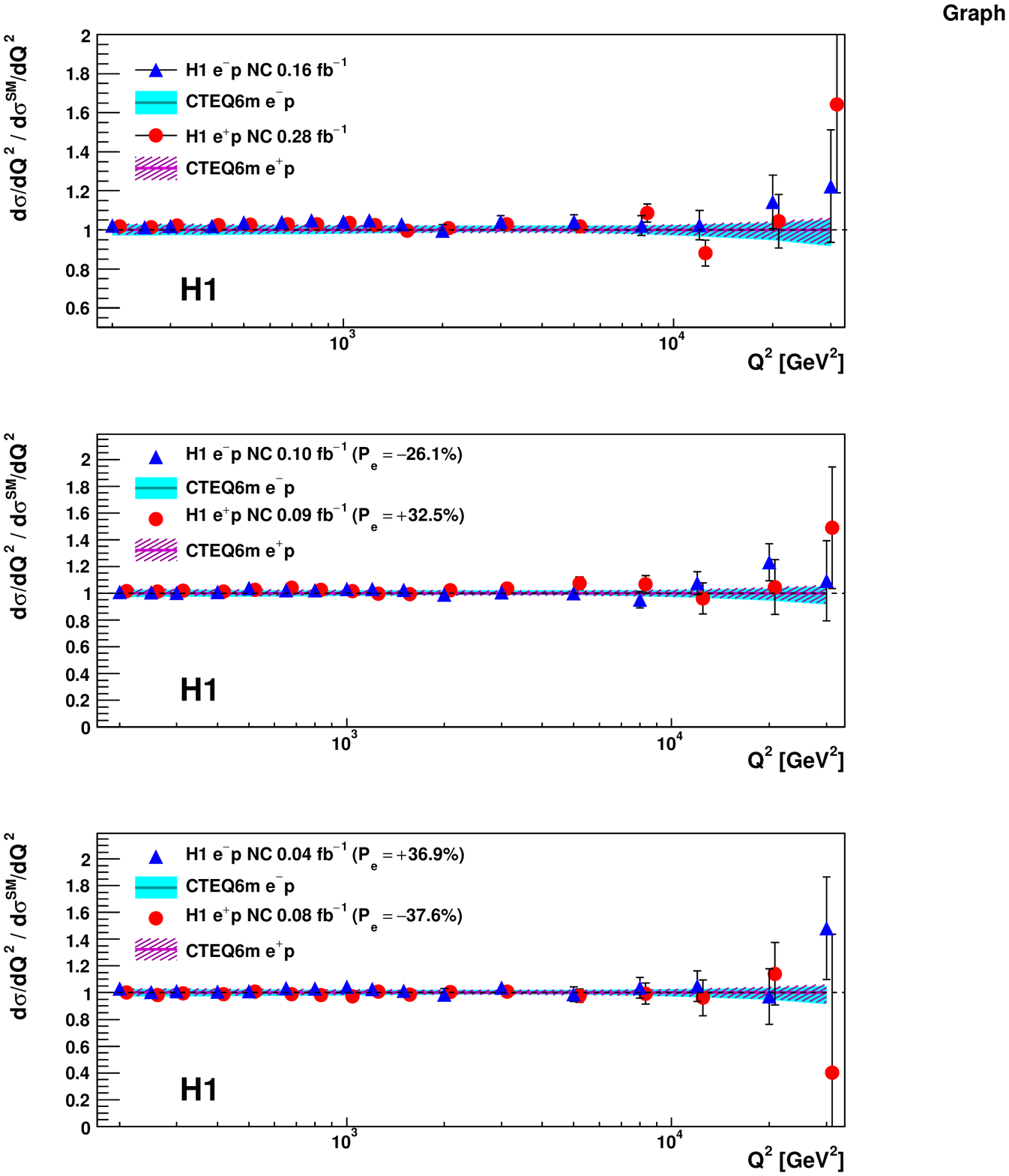,width=0.5\textwidth,clip=}
\hspace*{0.8cm}
\epsfig{file=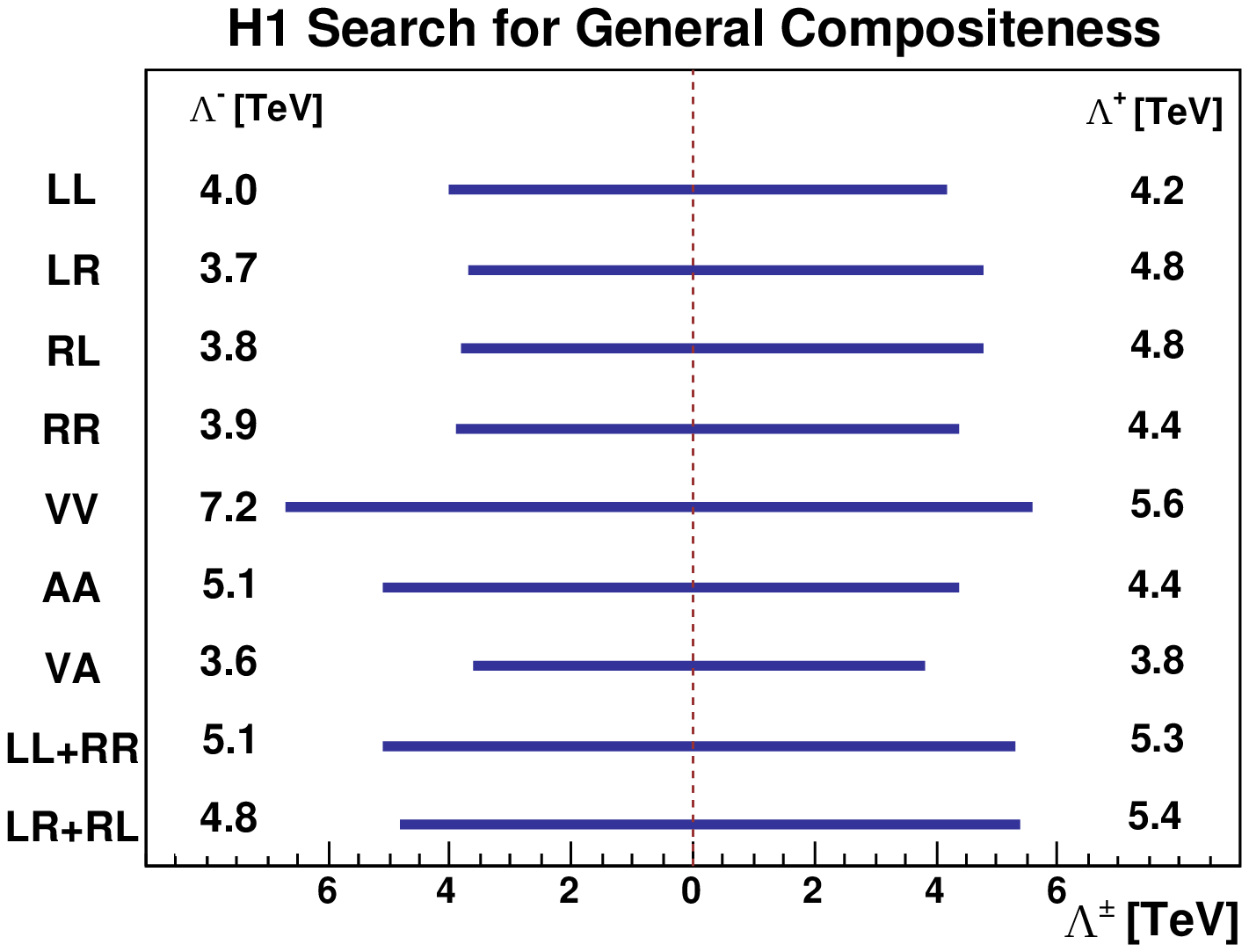,width=0.45\textwidth}
  \caption{ 
Left: 
The ratio of the measured cross section to the Standard Model
prediction, determined using the CTEQ6m PDF set, for polarised H1 NC DIS
data taken from the year 2003 onward for different lepton charge and
polarisation data sets. 
Right:  
Lower limits at 95\% CL on the compositeness scale $\Lambda$ for
various chiral models, obtained from the full H1 data. Limits are
given for both signs $\Lambda^-$ and $\Lambda^+$ of the chiral coefficients.
    \label{fig:h1ci} }
\end{figure}

DIS cross sections at HERA are also sensitive to possible new
interactions between electrons and quarks   
involving mass scales above the center-of-mass energy, which could
modify the cross sections at high $Q^2$ via virtual effects. This would
result in observable deviations from the Standard Model predictions. 
Many new interactions, such as processes mediated by heavy leptoquarks,
can be modeled as four-fermion contact interactions. 
The H1 Collaboration applied a common method to search for deviations from
SM predictions for different new physics scenarios, which can be
considered within this framework.
Data on scattering of polarised electrons and positrons collected 
from HERA~II were combined with electron and positron data from HERA~I.
No significant deviation from the Standard Model predictions was
observed, as shown in Figure~\ref{fig:h1ci} (left)
and 95\% limits were derived for the relevant parameters of the models
studied. 
For the general contact-interaction models, limits on the compositeness
scale, $\Lambda$, ranging from 3.6 to
7.2~TeV were obtained, as shown in Figure~\ref{fig:h1ci} (right).
The study of leptoquark exchange yielded lower limits on the ratio
$M_{LQ}/\lambda$ between 0.41 and 1.86 TeV. 
For models with large extra dimensions, scales below
0.96~TeV were excluded.
Finally, a quark-charge radius larger than $0.65\cdot 10^{-16}$ cm was
excluded, using the classical form-factor approximation.


\section{Electroweak cross sections}

\begin{figure}[t]
\begin{center} 
  \epsfig{file=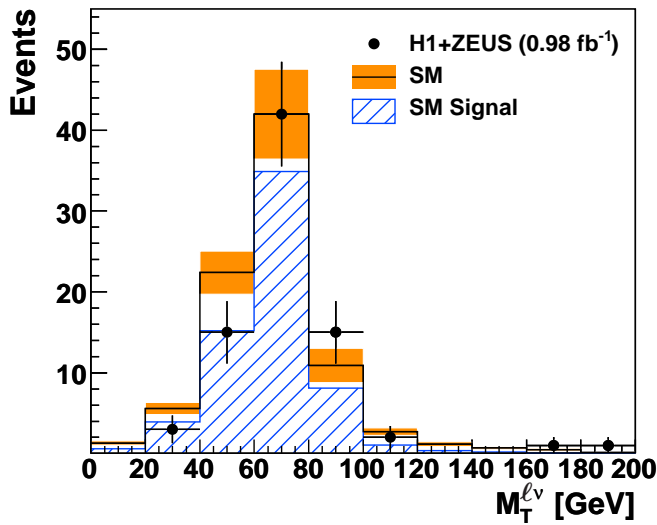,width=0.6\textwidth}  
\end{center}
  \caption{
Distribution of the lepton-neutrino transverse mass $M_{T}^{l\nu}$ 
for events with an isolated electron or muon and missing transverse 
momentum in the full HERA $e^{\pm}p$ data sample.
    \label{fig:h1zeus_w} }
\end{figure}

Both the H1 and ZEUS Collaborations have used the high statistics 
of the collected data to study a variety of processes sensitive to 
electroweak interactions, in particular the production of heavy weak bosons 
and the top quark.
One of the interesting signatures, which was considered in the
combined H1 and ZEUS analysis, is production of isolated high-$p_T$ leptons
(electrons or muons) and a large missing transverse momentum. 
The main SM process that may produce events with this topology is the
production of real W bosons via photoproduction with a subsequent 
leptonic decay: $ep \rightarrow eW^\pm (\rightarrow l\nu )X$. 
Only with the full HERA high energy data, corresponding to an
integrated luminosity of 0.98~fb$^{-1}$, a cross section measurement for
$W^{\pm}$ production in this process became accessible.
The expected numbers of signal and background events, after the final
selection cuts, were 64.7$\pm$9.9 and 23.1$\pm$3.3 respectively (total
of 87.8$\pm$11.0) and observed in the final data sample were 81 events. 
The distribution of the lepton-neutrino transverse mass, $M_{T}^{l\nu}$, 
for the final sample of selected H1 and ZEUS events is shown in
Figure~\ref{fig:h1zeus_w}. 
The resulting cross section estimate is:
\begin{eqnarray*}
\sigma_{W} & = & 1.06 \pm 0.16 (stat.) \pm 0.07 (sys.)\; pb \; ,
\end{eqnarray*}
in agreement with the SM prediction of
$  1.26\pm 0.19\; pb$.

\begin{figure}[t]
\begin{center} 
  \epsfig{file=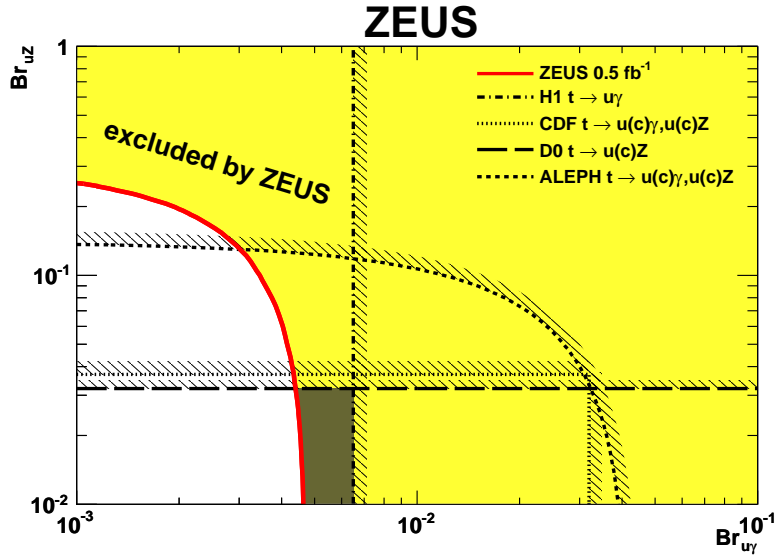,width=0.7\textwidth}  
\end{center}
  \caption{
ZEUS limits on the anomalous top quark couplings shown in the BR$(t
\rightarrow u\gamma)$  vs BR($t \rightarrow u Z$) plane. Also shown
are corresponding boundaries from H1, CDF, D0 and ALEPH
experiments. The shaded area is excluded. The dark shaded region
denotes the area uniquely excluded by ZEUS. 
    \label{fig:zeus_t} }
\end{figure}

A similar selection procedure can be used to search for single-top
production at HERA. 
In addition to an isolated lepton and missing transverse 
momentum, a large hadronic transverse momentum, $P_T^{had}$, is expected.
A $b$-tagging algorithm can additionally be used to suppress the $W$ 
production background. 
Even though the SM cross section for single top production at HERA is
below 1~fb, the measurement of this process is important,
as a significant enhancement is expected in several BSM scenarios
due to FCNC couplings.
A search for anomalous single top production was performed using the
data collected with the ZEUS detector and a limit on the production cross
section $\sigma < 0.13$ pb (95\% CL) was set\cite{desy-11-204}.
The cross section limits were converted into limits on the anomalous top
quark couplings and the branching ratios $t \rightarrow u
\gamma$ and $t \rightarrow u Z$, as shown in Figure~\ref{fig:zeus_t}.

\begin{figure}[t]
\begin{center} 
  \epsfig{file=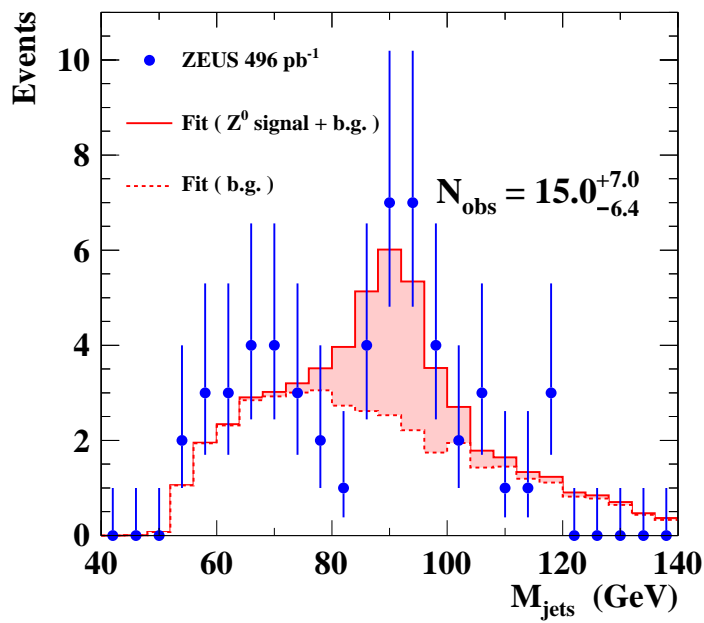,width=0.6\textwidth}  
\end{center}
  \caption{
The invariant-mass distribution for the selected sample of elastic and
quasi-elastic $Z^{0}$ production events. ZEUS data (solid points) are
compared with the fit result for the signal plus background and the
background shape template based on the inelastic events.
    \label{fig:zeus_z0} }
\end{figure}

Also in case of the $Z^{0}$ production at HERA, the SM cross section is
predicted to be very low, about 0.4~pb.
The ZEUS Collaboration studied this process using an integrated luminosity
of about 0.5 fb$^{-1}$.
Thanks to the excellent energy resolution of the
ZEUS hadronic calorimeter, the hadronic decay mode could be used in this
analysis, significantly increasing the expected event rate.
The analysis\cite{desy-12-168} was restricted to elastic and
quasi-elastic $Z^{0}$ production in order to suppress QCD multi-jet
background.  
Figure~\ref{fig:zeus_z0} shows the invariant-mass distribution of the
selected events. It also shows the fit result for the signal plus
background and the background separately.
The invariant mass distribution measured for inelastic events was used as
a background template in the fit.
The fitted number of observed $Z^{0}$ events is 15.0$^{+7.0}_{-6.4}$,
which corresponds to a $2.3\;\sigma$ statistical significance.
The cross section for the elastic and quasi-elastic production of
$Z^{0}$ bosons at $\sqrt{s} = 318$~GeV
was calculated to be
\begin{eqnarray*}
\sigma(ep\rightarrow ep^{(*)}Z^{0}) & = & 
   0.13 \pm 0.06 (stat.) \pm 0.01 (syst.)\; pb \; .
\end{eqnarray*}
This result is consistent with the SM cross section estimate of
0.16~pb.


\section{Conclusions}

With high luminosity and lepton beam polarisation, HERA provided
a unique window for precise electroweak studies. 
Although data taking was completed in 2007, 
the H1 and ZEUS collaborations are still working hard, 
making progress in understanding the detector and finalizing various 
data analyses. 
The results presented at this conference are only a small selection of
recently completed work.

The presented results on the NC and CC DIS at high $Q^{2}$, including charge and
polarisation asymmetries, are in very good agreement with the SM.
With the high precision and the large kinematic coverage of the data, 
the NLO QCD analysis was extended to extract not only parton densities in
the proton but to fit electroweak parameters as well. 
The obtained constraints on the light-quark couplings to the $Z^0$ boson 
are in good agreement with the Standard Model predictions and are
competitive in precision with LEP and Tevatron measurements. 
The precise measurements of deep inelastic $e^\pm p$ scattering at large
$Q^2$ were also exploited to search for possible ``new physics'' beyond
the Standard Model.
As no significant deviation from the Standard Model predictions was observed,
limits were derived for different models of new physics.
     
The production of electroweak bosons and top quarks in $ep$
collisions are also good benchmark processes for testing 
the Standard Model.
The full HERA data sample from both experiments was analysed,
corresponding to a total integrated luminosity of 0.98~fb$^{-1}$, in
a search for $W^{\pm}$ production. The total and differential
single $W$ production cross sections were measured to be in agreement
with the SM predictions. 
An analysis of the ZEUS Collaboration resulted in the first observation
of $Z^{0}$ production in $ep$ collisions, with  15.0$^{+7.0}_{-6.4}$
signal events, corresponding to a $2.3\;\sigma$ statistical significance. 
The resulting cross section for elastic and quasi-elastic production of
$Z^{0}$ bosons is consistent with the SM prediction.


\end{document}